\documentclass[twocolumn, amsmath, superscriptaddress, amsfonts,prb]{revtex4-2}
\usepackage{graphicx}
\usepackage{url}
\pdfoutput=1
\usepackage{subfigure}
\usepackage{mathtools}
\usepackage{float}
\usepackage{lipsum}

\begin{document}

\title{Radiation enhanced diffusion in cartilages as a physical mechanism underlying radiation treatments of osteoarthritis and related disorders}

\author{Diana Shvydka}\email{diana.shvydka@osumc.edu}\affiliation{Dept. of Radiation Oncology, The Ohio State University Wexner Medical Center, Columbus, OH 43221, USA}
\author{Victor Karpov}\email{victor.karpov@utoledo.edu}\affiliation{Department of Physics and Astronomy, University of Toledo, Toledo,OH 43606, USA}

\begin{abstract}
Degradation of joint cartilages can result in osteoarthritis (OA) affecting about 10\% of the US population and responsible for significant hospitalization costs. While observations show that low dose radiation treatments (LDRT) bring improvements for a majority of OA patients, the underlying mechanism is not sufficiently understood. Here, we show how the radiation enhanced diffusion (RED) can boost the molecular transport in cartilages promoting cartilage self-healing rendering a mechanism for the observed positive LDRT effects on OA. Along with quantitative estimates for RED, we predict a related phenomenon of the electric charge build up that allows LDRT schedules promoting desirable types of molecular transports dominated by either positive or negative molecular species. Our analyses call upon further experimental verifications and clinical trials with curative rather than palliative intent. In addition to OA applications, our developed approaches can be useful for sports medicine dealing with damage or degeneration of the articular cartilages.

\end{abstract}

\maketitle

\section{Introduction}\label{sec:intro}

Osteoarthritis (OA) is a progressive disorder with signs of joint stiffness, pain, and loss of mobility in joints of hands, hips, and knees affecting over 30 million Americans; it is estimated that by the year 2040, over 10\% of all adults will experience OA-related symptoms. OA is the second-most costly health condition in the US and is responsible for over 4\% of all total hospitalization costs.

Low dose radiation treatments (LDRT) [typical doses of 0.3 – 0.7 Gy a day, with 1 Gy daily limit to avoid secondary malignancy] has been shown to provide improvements in 60\% to 90\% of patients (for recent clinical reviews and treatment protocols see \cite{weissman2023,donau2020}). LDRT for OA is minimally used in the US despite being a cost-effective noninvasive treatment with minimal side effects. Less than 10\% of providers in the US use LDRT for OA while over 85\% of providers in European countries do.

OA results from the degeneration of articular cartilage between bones in the joint, although the understanding of its underlying mechanism remains insufficient. That degradation is not capable of self-healing due to lack of efficient transport in cartilages. Indeed, cartilage is an avascular and alymphatic tissue, in which the primary mode of transport for nutrients, oxygen, waste products, signaling molecules, and matrix macromolecules is by diffusion. Therefore, it is important to understanding the factors that can influence diffusive transport in cartilages.

In what follows, we emphasize the role of radiation enhanced diffusion (RED) as a possible mechanism behind the observed positive effects of LDRT on OA. We show that RED can amplify molecular diffusivity by many orders of magnitude depending on the radiation type, intensity, and material parameters. In spite of its possible strong effects known for semiconductor and other materials, RED has not been considered with living tissues, particulary with OA LDRT. 

Furthermore, we identify in what follows another important effect of ionizing radiation in cartilages as materials with very low electronic mobilities leading to the electric charge build-up. That build-up created long-lived nonuniform distributions of electric fields that promote molecular transport. We will argue that depending on the LDRT details, the field accelerated transport can favor any desired direction and ionic polarity. This opens a venue to purposely tailor the polarity and type of molecular transport achieving the favorable balances between the positive and negative ions beneficial for the cartilage structure.

Despite the existing optimistic views of the LDRT effects, \cite{dove2022,weissman2023,ruhle2021,alvarez2020} some studies have observed no or negative effects of radiation on cartilages, often at doses significantly higher than those recommended in OA LDRT. \cite{willey2013,willey2016,saintingy2015} While these studies have drawn different conclusions on the effects of ionizing radiation on articular cartilage, they support the same overall verdict that OA LDRT research needs to be continued and broadened in order to make a consistent conclusion. \cite{cash2019}

While focused on basic physical phenomena, our approach here does not neglect the previously discussed radiobiological mechanisms of LDRT OA effects, such as anti-inflammatory efficacy based on the modulation of a multitude of inflammatory pathways and cellular components, including endothelial cells, leukocytes, and macrophages as reviewed in Ref. \onlinecite{dove2022}. Rather, our approach presents a complementary view aimed at explaining the basic physics underlying the radiobiological mechanisms of OA LDRT. More specifically, we discuss the LDRT related molecular transport modes logically related to the radiobiological mechanisms.

\begin{figure*}[t!]
\hspace{-1.3cm}\includegraphics[scale = 0.37]{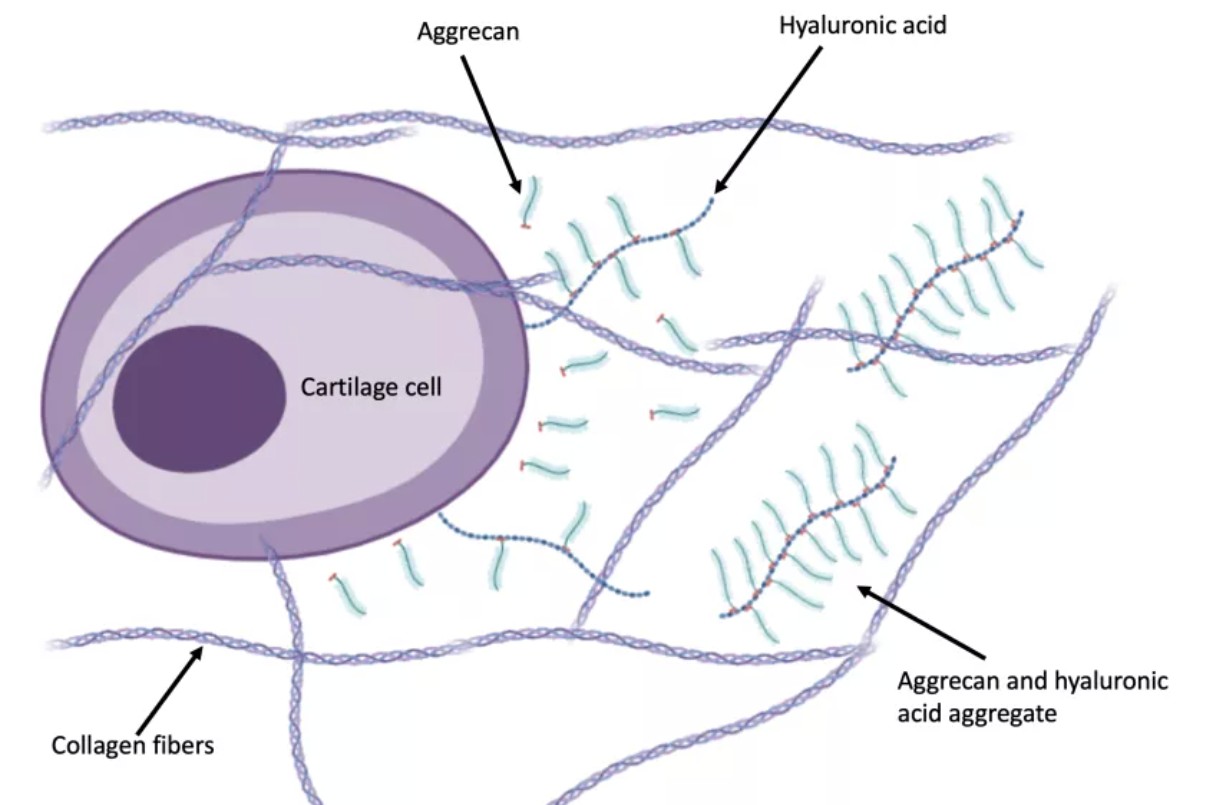} \hspace{1.0cm}            \includegraphics[height=0.28\textwidth,]{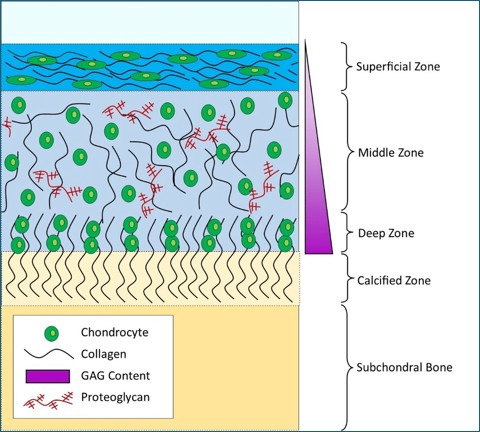}
\caption{Schematics of a cartilage structure. Left: A cartilage cell is surrounded by extracellular matrix consisting of proteins (collagen fibers), non-proteoglycan polysaccharides (hyaluronic acid), and proteoglycan (aggrecan). Illustrations downloaded from: Biorender.com . Right: The structure of articular cartilage and the underlying subchondral bone. From articular cartilage surface to the bone there are four zones—superficial, middle, deep, and calcified. Within the zones, there are differences in the orientation of the collagen fibers, the arrangement of the chondrocytes, and the distribution of proteoglycans and their associated GAGs. Downloaded with permission from Ref. \onlinecite{davies2019}.}\label{Fig:cartilagecell}
\end{figure*}

This manuscript is organized as follows. Section \ref{sec:surv} gives a brief survey of cartilage structures including the OA related changes. The established mechanisms of molecular diffusion in cartilages are described in Section \ref{sec:cartdif}. RED mechanisms and their estimates are described in Section \ref{sec:RED}. A concomitant phenomenon of the electric charge build-up is introduced in Section \ref{sec:chargebuild} where we also describe its effects on molecular transport and various LDRT planning types. Finally, Section \ref{sec:concl} will summarize our  conclusions.

\section{Brief survey of cartilage structures}\label{sec:surv}
The schematic structure of articular cartilages in unspecified joints is illustrated in Fig. \ref{Fig:cartilagecell}. Their characteristic thicknesses are in the region of $\sim 0.5-3$ mm. The articular cartilages consist of cartilage cells (also known as chondrocytes) surrounded by extracellular matrix (ECM). The ECM throughout a cartilage is composed of water, collagen, and proteoglycans, as well as some non-collagenous proteins and glycoproteins in lesser amounts with a distribution of chondrocytes. Other important molecules, glycosaminoglycans (GAGs), are large, negatively-charged polysaccharides made up of repeating disaccharide units. In addition, there are the glycoproteins and proteoglycans, which are complex macromolecules made up of proteins covalently linked to carbohydrates. \cite{fox2009,cornellisen1996}

Basically, ECM is a network of proteins and sugar molecules containing a large proportion of water, which helps creating an impact-absorbing effect in the joint and reduces joint friction. As it is the cartilage cells that produce and maintain ECM, healthy cartilage cells are needed for the cartilage to function normally.

In the ECM, a network of the structural protein collagen works as the cartilage's "armour". Inside this "armour” lies the protein aggrecan, whose role is to bind water and keep the cartilage soft and elastic so that it swells and becomes a ”cushion" in the joint. The collagen prevents the aggrecan from absorbing too much water and swelling excessively. The thickness of the cartilage is mainly dependent on the interaction between collagen and aggrecan.

A positively charged water hydrogen is attracted to the GAGs. Proteoglycans are bound to each other forming large negatively-charged aggregates. Their concentration increases towards the bone. Although a cartilage is avascular, aneural and alymphatic, its chondrocytes must receive nutrients and oxygen to maintain tissue function: synthesis of collagen, proteoglycans and glycosaminoglycans.

The cartilage stiffness and elasticity are attributed to ECM swelling due to  GAGs-water attraction. The concomitant movement of interstitial fluid along with frictional drag forces, gives rise to the time-dependent poroelastic behavior responding to compressive loads and strain.

In general, the degradation of articular cartilage in OA is the result of imbalance in the construction and degradation of ECM. This may happen when the cartilage cells do not build up ECM at the same rate it is broken down. This gradual degradation of ECM creates cartilage damage. There is no exact answer as to why the above imbalance occurs it is generally related to either oblique loading, overloading, or underloading of the joint. \cite{goldring2012}

During the development of OA, the chondrocytes undergo a phenotypic shift, resulting in fibrillation and degradation of
ECM, the appearance of chondrocyte clusters, increased cartilage calcification associated with tidemark advancement or duplication, and vascular penetration from
the subchondral bone. Coincident matrix degradation products can further promote catabolic activation, aberrant, hypertrophy-like differentiation, and apoptosis.
Once the collagen 'armour' is degraded, it cannot be repaired to its original state for the lack of sufficient transport. \cite{goldring2012}

In the first stage of degradation, the aggrecan content in the cartilage decreases while the collagen network remains intact and so does the cartilage thickness. During the second stage, the collagen network starts breaking down and the amount of aggrecan keep decreasing. As a result, less water is absorbed and the cartilage starts thinning,its surface becoming soft and uneven. Over longer times, perhaps in the course of years, the cartilage in some patients disappears completely. Overall, the degradation of the cartilage appears slow making OA a degenerative disease, which however is not reducible to wear and tear processes.\cite{davies2019}

Overall, the problems with articular cartilage seem to be related to its very limited capacity for self repair, where mall damages only further deteriorate over time. In the absence of sufficient vasculature regenerating damaged cartilages would require significant enhancement in molecular transport mechanisms other than blood flow such as anomalously strong diffusion. Should the efficient transport be achieved, it could  supply the necessary degradation protecting and regenerating molecular components, \cite{hou2021} and remove from the cartilage its  deteriorating products. \cite{yasuda2006,goldring2012,karsdal2008}


\section{Thermal diffusion in cartilages}\label{sec:cartdif}
Alterations in the rates of molecular diffusion, which may occur with aging or disease, have the potential to influence chondrocyte metabolism
by affecting the transport of nutrients, signaling molecules, or pharmacologic agent in cartilages. Studies of molecular diffusion in cartilages included various species, cartilage domains, loads, etc. \cite{leddy2003,leddy2007,lee2011,raya2015,meng2020,travascio2020} The experiments provide broad evidence that diffusion is sensitive to the protoglecan loss, that the collagen network is responsible for the diffusion
anisotropy, and that diffusion parameters correlate with the mechanical properties of cartilage.

The diffusion of neutral solutes through cartilages is affected by the loading patterns. For example, OA cartilage in
early and middle stages, accelerated contrast agent infiltration is found  under cyclic loading compared with static loading.
However, for late-stage OA samples, no acceleration of diffusion was observed in the first hours because of the
insufficient resilience of compressed cartilage. The accumulation of neutral solutes in an upward invasive fissure
also suggested that solutes could penetrate into the fissure under cyclic loading.\cite{meng2020}

The rate of molecular diffusion in articular cartilage exhibits specific effects that often depend on the size and nature of the diffusing molecule. For example, diffusion coefficients in cartilage are generally inversely proportional to molecular size $R$ \cite{torzilli1987,leddy2003} and increase with temperature $T$ qualitatively in accordance with the Stokes–Einstein relationship.\cite{torzilli1993}
Similar to many other leaving tissues, the interval of temperatures cannot be made broad enough for more quantitative conclusions. Therefore, the authors refer to the Stokes- Einstein model, in which the diffusing particle is spherical and the diffusion coefficient $D$ is presented as \cite{kittel}
\begin{equation}D_\eta =k_BT/6\pi R\eta \label{eq:SE}\end{equation}
where $k_B$ is the Boltzmann's constant and $\eta$ is the viscosity.

In solids and complex liquids, $\eta$ is often determined by processes of thermal activation (related to molecular bonds restructuring), which is natural to describe as
\begin{equation}\eta =\eta _0\exp (W_B/k_BT)\label{eq:eta}\end{equation}
where $W_B$ is the activation barrier and $\eta _0$ is a constant. Indeed, that type of activation was observed in many complex fluids. \cite{seeton2006} Therefore, a blind reference to the Stokes-Einstein formula of Eq. (\ref{eq:SE}) does not really mean the proportionality to $T$ pointing rather at a trend that could be attributed to the exponential dependence of Eq. (\ref{eq:eta}). Neither that reference implies a spherical shape of a diffusant; instead, $R$ should be interpreted as its characteristic linear dimension. Structural transformations in the process of diffusion of an unspherical molecule appear conceivable.

Should the latter be the case, a particular component of simpler structure, such as water, would not contribute significantly to the measured diffusivities in complex fluids. Indeed, water content across the cartilage structure did not strongly affect the measured diffusivity. In contrast, significant correlations were found among diffusivities of large molecules and GAG content. \cite{travascio2020} The diffusion anisotropy in cartilages was found to be not very significant (tens of percent). \cite{raya2015} Similarly, the diffusion coefficients increase with cartilage degradation remaining within the same order of magnitude.

\begin{figure*}[hbt!]
\hspace{-1.3cm}\includegraphics[scale = 0.52]{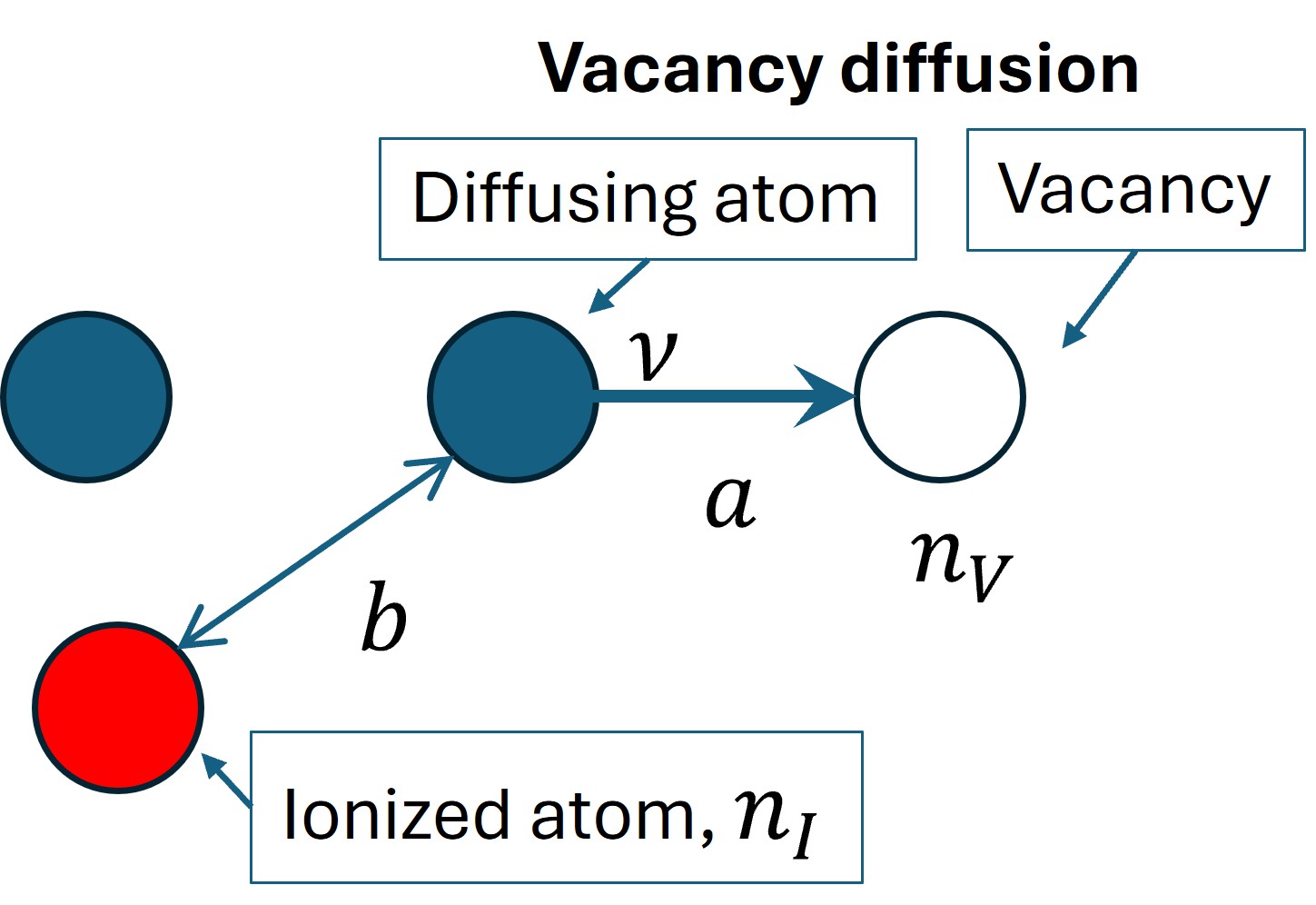} \hspace{1.5cm}            \includegraphics[scale=0.54]{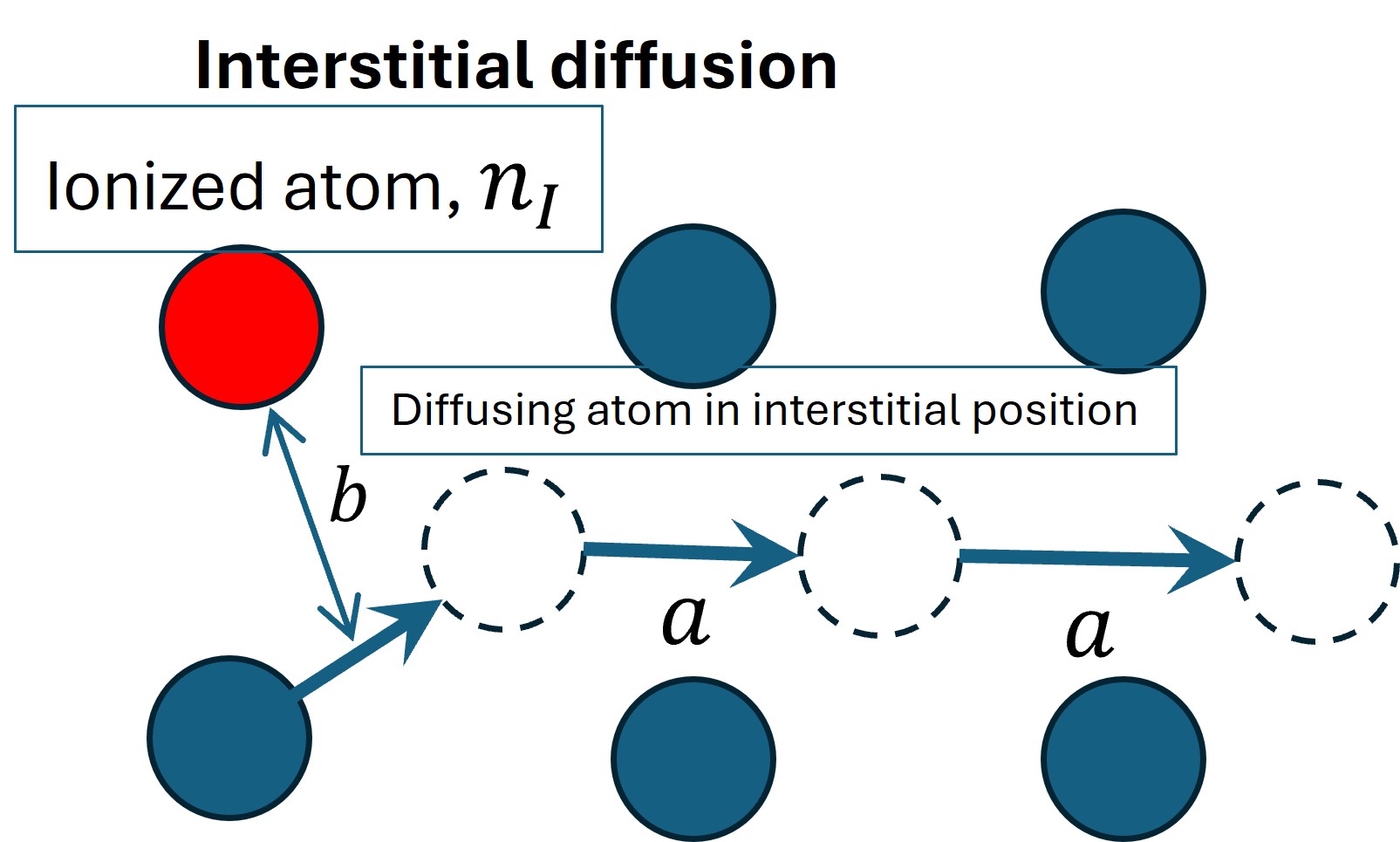}
\caption{Sketches of two RED microscopic mechanisms. Left: vacancy mediated diffusion where an ionized atom creates Coulomb repulsion moving nearby ions to vacancy cites. $n_I$ and $n_V$ represent respectively the concentrations of ionized atoms and vacancies. Right: diffusion of ions (dashed circles) through the intersticial positions. The diffusing ions are moved by the electric forces due to a radiation ionized atoms distance $b$ away. }\label{Fig:RED}
\end{figure*}

The same can be stated about the position dependent diffusion that may be affected by the local composition
and structure, which vary with depth from the tissue surface. While the diffusion
coefficients for various molecules vary significantly with depth, these variations between deep and surface  zones remain within one order of magnitude,
reflecting variations in the structure and composition of collagen, proteoglycans, and other macromolecules among the zones. \cite{leddy2003}

Phenomenologically it is natural to describe molecular diffusion in cartilages in the same framework as diffusion in other condensed matter, as a sequence of thermoactivated atomic displacements of length $a$  each. Describing the frequency of thermoactivated events by the Boltzmann's exponent, the diffusion coefficient is estimated as, \cite{cussler2009,boltaks2021}
\begin{equation} \label{eq:D}
D=n_V\nu _0a^2\exp(-W_B/k_BT).
\end{equation}
Here $n_V$ is the relative concentration of vacant sites and $\nu _0\sim 10^{13}$ s$^{-1}$ is the frequency of attempts, of the order of the characteristic atomic vibrational frequency in condensed matter. The preexponential of Eq. (\ref{eq:D}) is supposed to include in some form the quantity $D_\eta /a^2$ from Eq. (\ref{eq:SE}),  which can be treated as a constant compared to the exponential term $\exp(-W_B/kT)$.

From Eq. \ref{eq:D}, one can estimate $W_B$ from the typical measured values \cite{leddy2007,meng2020,maroudas1970,ren2023} $D\sim 1-100 $ $\mu$m$^2$s$^{-1}$. Assuming $a\sim 1$ nm, $T=300$ K, and $n_V\sim 1$ yields
\begin{equation}\label{eq:WB}W_B=k_BT\ln(n_V\nu_0 a^2/D)\sim 0.5-1 {\rm eV}, \end{equation}
which is in the ballpark of diffusion barriers in condensed matter \cite{cussler2009,boltaks2021} and complex fluids (when approximated by Eq. (\ref{eq:WB}). \cite{seeton2006}

Based on the above mentioned numerical values for $D$, one can compare the potential supply of nutrients via diffusion to that by direct blood flow, which can be the building material for self-healing processes in living tissues. We denote by $c$ the relative concentration of a particular chemical component of blood and use the characteristic (drift) blood velocity of $v\sim 10^2$ cm/s. Then the flux density of that component is estimated as
\begin{equation}\label{eq:flux}\phi _{\rm drift} = cnv\sim c\times 10^{14} \quad {\rm cm}^{-2}{\rm s}^{-1}\end{equation}
where $n\sim 10^{12}$ cm$^{-3}$ is the characteristic concentration of blood cells.

\begin{figure*}[htb!]
\hspace{-1.3cm}\includegraphics[scale = 0.45]{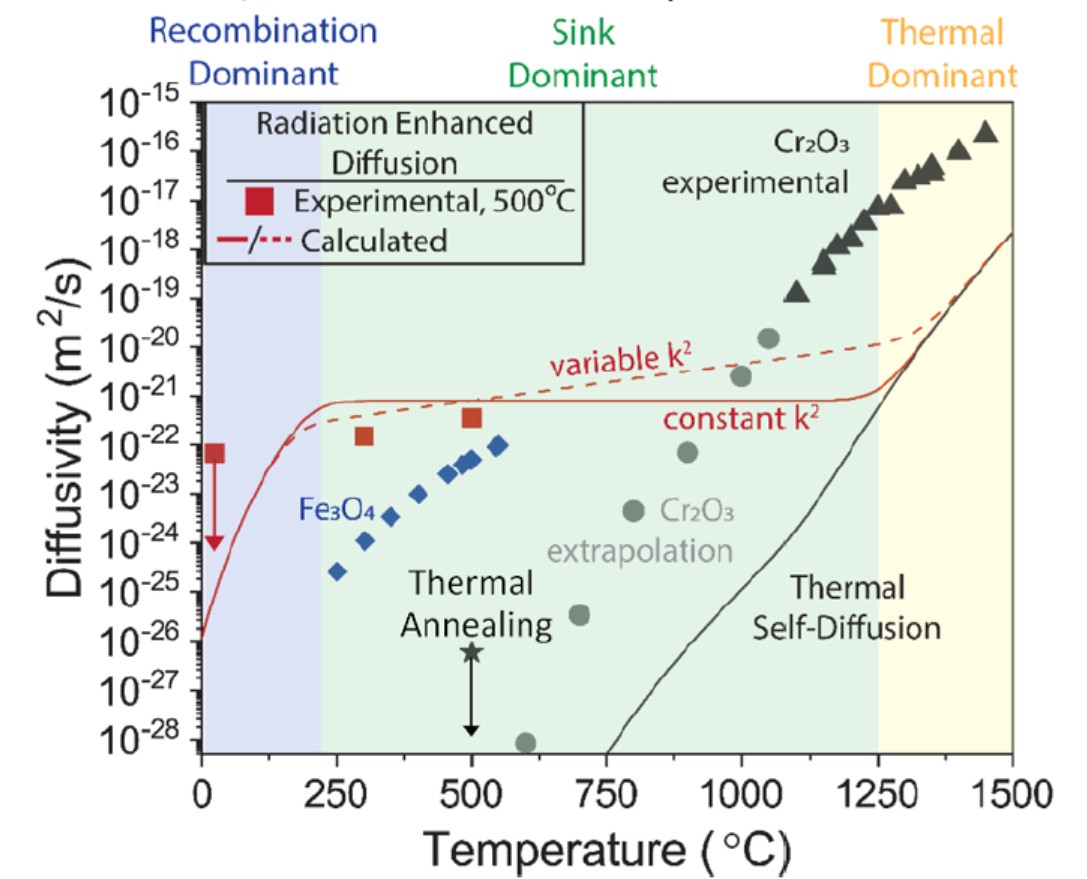} \hspace{1.1cm}            \includegraphics[scale=0.34]{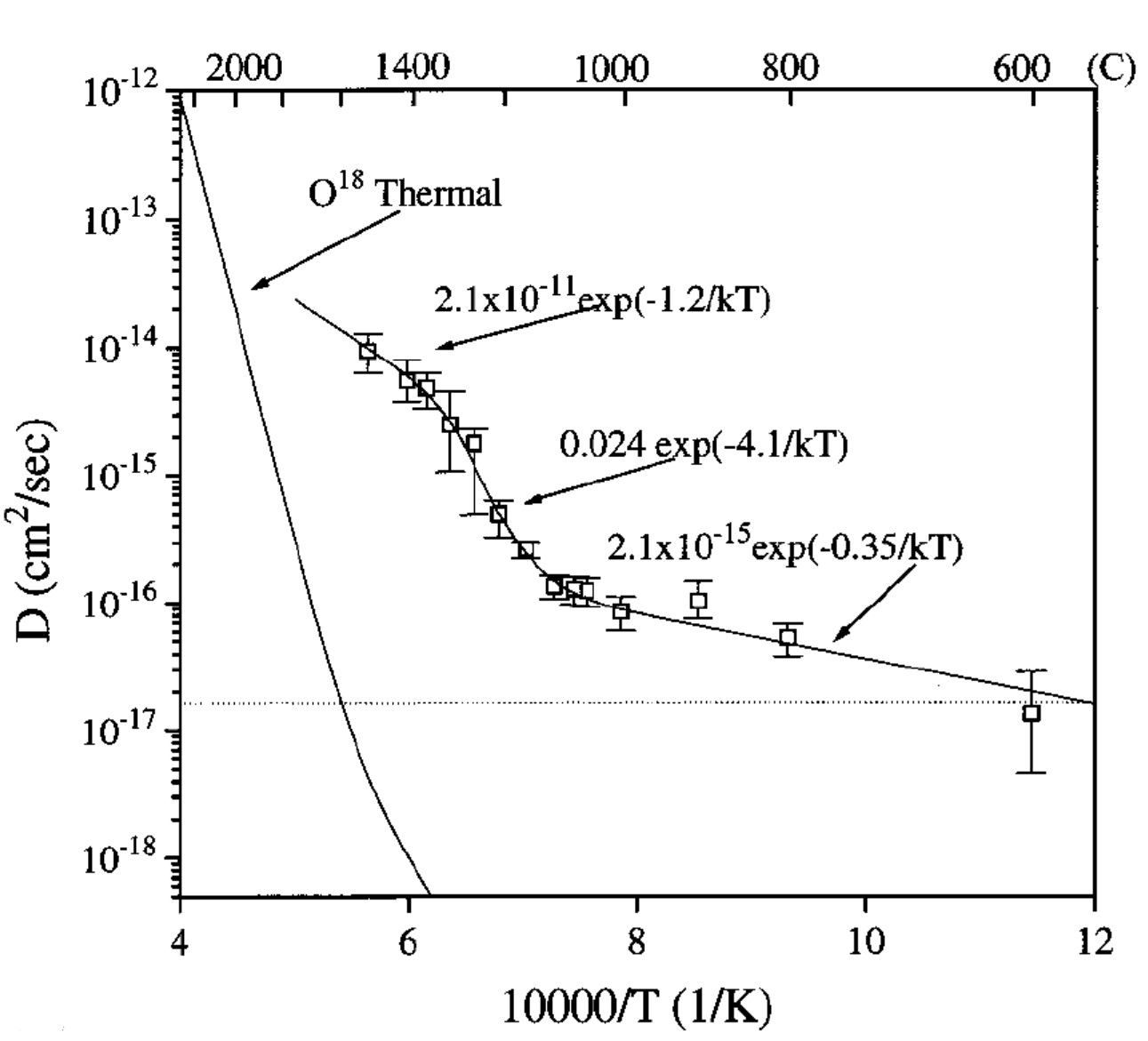}
\caption{Representative data on RED. Left: Experimental diffusivity is compared to literature results for Cr$_2$O$_3$ \cite{hagel1965}
and Fe$_3$O$_4$.\cite{castle1969} Overlaid on the results is the predicted self-diffusion of oxygen by the chemical rate-theory model. Here the parameter $k^2$ represents the effect of vacancy sinks, beyond the current scope. The dose rate is described as $2\times 10^{-4}$ dpa/s. Courtesy of ref. \cite{yano2021}. Right:  Radiation-enhanced tracer diffusion coefficients in MgO during 2.0 MeV Kr irradiation. The thermal diffusivity is indicated for
comparison. (Reproduced from Ref. \cite{sambeek1998} with the permission of AIP Publishing.) See also data from Ref. \cite{chen2020} where coefficients for nickel self-ion irradiation at steady state as functions of reverse temperature with damage rates of $10^{-2}$, $10^{-3}$, and $10^{-4}$ displacements per atom per second (dpa/s). }\label{Fig:REDATA}
\end{figure*}

To estimate the diffusion driven flux of the same component we assume that its concentration diminishes from $cn$ to zero across the characteristic cartilage thickness  $l\sim 1$ mm. This yields the diffusion flux density estimate,
\begin{equation}
\phi _{\rm diff}=D|grad (cn)|\sim \frac{Dnc}{l}.\label{eq:diflux}\end{equation}

The ratio of the two,
\begin{equation}\alpha=\frac{\phi _{\rm diff}}{\phi _{\rm drift}}\sim \frac{D}{lv}\sim 10^{-9}-10^{-5}.\label{eq:ratio}\end{equation}
describes the relative slowness of diffusion vs. drift dominated self-healing. Assuming for example the characteristic drift-dominated self-healing time $t _{\rm drift} \sim 10$ days (such as for a minor finger cut), the diffusion dominated self-healing of a cartilage would take from 1 million to 10 billion days. While the conclusion of diffusion being inefficient self-healing mechanism has long been known,  the above simplistic estimates present it more quantitatively.

In summary, the thermal diffusion in cartilages appears to be as one would generally expect for complex structures: somewhat anisotropic, somewhat position dependent, and characterized by the diffusion barriers of about 'normal' height. It is not surprising then that such a diffusion in an avascular cartilage tissue cannot make up for the much more efficient convection transport typical of various vascular tissues; hence, lack of self-healing. We show next how radiation can enhance the molecular diffusion in cartilages by many orders of magnitude.

\section{Radiation enhanced diffusion}\label{sec:RED}

Radiation-enhanced diffusion (RED) is a phenomenon, wherein the presence of radiation accelerates the diffusion of atomic particles in materials. The RED arises because of the creation of defects in the material lattice, such as vacancies or interstitials, or altering the ionic charges by the radiation. The phenomenon of RED was identified long time ago \cite{dienes1958} and then extensively studied through the years (see e. g. Refs. \onlinecite{dienes1958,kaschny1993,starostin2005,doyle2018,yano2021,chen2020,zinkle1997,peak1976,bourgoin1978} to name a few). Note that some publication discriminate between the general term of RED and more specific 'radiation induced diffusion' or `ionization induced diffusion'.\cite{zinkle1997,peak1976,bourgoin1978} Various microscopic interpretation \cite{zinkle1997,bourgoin1978} predict diffusion coefficients varying by orders of magnitude, and yet difficult to verify experimentally.

Shown in Fig. \ref{Fig:RED} are sketches of two conceivable mechanisms of RED; others can be proposed along similar lines. \cite{kaschny1993,starostin2005,doyle2018,yano2021} The vacancy diffusion mechanism implies molecular movements assisted by the neighboring vacated sites providing enough room for the displacements. The role of radiation here is twofold: (i) it can create more vacancies and intersticials (of concentration $n_V=n_I$), and (ii) it can alter the charge of neighboring atoms (short distance $b$ apart) thus changing the diffusion barrier by the Coulomb interaction. Another part of Fig. \ref{Fig:RED} shows RED through  the intersticial pathway, not qualitatively differen from the vacancy mechanism. For brevity, we consider here the vacancy dominated RED trying to present conceivable estimates of RED enhancements.

We start with noting that a major mechanism of RED in the literature is related to the radiation induced increase in vacancy concentration due to the radiation caused displacements per atom per second (dpa/s). The relative concentration of vacancies $n_V=N_V/N_0$ is limited by their disappearance through recombination with interstitials at the rate $\gamma N_VN_I=\gamma N_0^2n_In_V=\gamma N_0^2n_V^2$ where $\gamma$ is the vacancy-interstitial recombination coefficient, or through absorbtion by the system sinks and external surfaces. While both channels are considered realistic, here we neglect the sink effects trying to minimize the number of independent parameters. Then, the relative vacancy concentration per atom is described by
\begin{equation}\label{eq:Nv}
\frac{dn_V}{dt}=\nu \exp\left(-W_V/k_BT\right) +I\sigma  -n_V^2N_0\gamma ,\end{equation}
where $t$ is time, $W_V$ is the vacancy formation energy, $I$ is the radiation intensity (per cm$^2$ per time), and$\sigma$ is the cross section of the process of atomic displacement by radiation.

As given by the thermodynamics \cite{kittel}, the quasistationary ($dn_V/dt=0$) equilibrium concentration of vacancies is
\begin{equation}\label{eq:vacterm}n_{V0}=\nu\exp(-W_V/2k_BT)/\gamma. \end{equation} Comparing with the latter, one can conclude that $$\nu /N_0\gamma \sim 1.$$ Therefore, Eq. (\ref{eq:Nv}) yields the quasistationary concentration of vacancies,
\begin{equation}\label{eq:NvI}  n_V=n_{V0}\sqrt{\left[1+\frac{I\sigma }{\nu}\exp\left(\frac{W_V}{k_BT}\right)\right]}.\end{equation}
We observe that
\begin{equation}\label{eq:figmer}\beta\equiv (I\sigma /\nu )\exp(W_V/k_BT)\end{equation}
is a figure of merit for the vacancy dominated RED. A similar consideration for the case when sink effects dominate predicts $n_V\propto I$ (rather than the above $\sqrt{I}$).

When $\beta\gg 1$, Eqs. (\ref{eq:vacterm}) and (\ref{eq:NvI}) predict the increase in vacancy concentration,
\begin{equation}\label{eq:deltanV} \frac{n_V-n_{V0}}{n_{V0}}\approx \sqrt{\frac{I\sigma}{\nu}}\exp \left(\frac{W_V}{k_BT}\right),\end{equation}
which we will semiquantitatively interpret as DPA. For numerical estimates, we use the typical radiation therapy linac's photon intensity \cite{yano2022} $I\sim 10^9$ cm$^{-2}$s$^{-1}$, $\sigma \sim 500$ barns \cite{oen1965} $\nu\sim 10^{13}$ s$^{-1}$, and $W_V$ of the order of several eV, Eq. (\ref{eq:deltanV})
 yields quantities in the range of $10^{-6} - 10^{-3}$.

It would be appropriate at this point to make a comparison with experimental data; however, no data on RED in cartilages or similar systems are available; hence, we resort to other systems RED data in Fig. \ref{Fig:REDATA}. The modeling results presented in Fig. \ref{Fig:REDATA} account for both the vacancy and intersticialcy mechanisms, and include the sink effects.

Keeping in mind Eq. (\ref{eq:D}) and  identifying $ n_V$ with dpa/s, we observe from both graphs in Fig. \ref{Fig:REDATA} gigantic, in excess of 10 orders of magnitude RED effects. This becomes clear when we linearly extrapolate the purely thermal diffusion coefficients down to room temperature ($T=300$ K) at which radiation can be applied to cartilages. Our conclusion here is consistent with the earlier published statement \cite{yano2022} that radiation enhanced diffusivities are found to be at least 20 orders of magnitude faster than the Arrhenius extrapolation of high temperature values.

We note as well that despite chemically differences between the systems (based on Cr, vs. oxides of Fe and Ni,) the order of magnitude effects are comparable given the same dose rate of $\sim 10^{-4}$ dpa/s. The above estimate  for $n_V$ predicts the result in the sane ballpark. The latter fact points at the possibility that similar RED figures can apply to cartilages. Such the RED enhancement scale readily overrides the insufficiency of thermal diffusion noted in Eq. (\ref{eq:ratio}) giving reasons to believe that RED can serve as a mechanism behind cartilage self-healing after LDRT.

That said, one should keep in mind that the rate of nutrient supply may be not the only important parameter determining the possibility of self-healing.  The rate of that nutrient 'digesting' will certainly have a decisive effect. It cannot be described solely in the realm of physics calling upon biological insights and leaving the question of RED induced self-healing open.

\section{Electric charge build-up}\label{sec:chargebuild}
One aspect typically overlooked in discussing radiation effects in living tissues is the electric charge build-up. Cartilages and many other tissues behave as insulating (or semi-insulating) materials; for example, the characteristic cartilage conductivity is $\sim 1$ S/m,  compared to $\sim 10^3$ S/m for  a classical semiconductor Si, and $\sim 10^9$ S/m for Cu.  The published work indicates the presence of strong electric fields inside and outside insulating films due to charging under electron beams of various energies. \cite{lai1989,melchinge1995,song1996,mizuhara2002,miyake2005,touzin2006,dennison2009,wilson2013,vasko2015} Overall, the sign of the charge acquired under electron irradiation depends on the material composition, interface and, electron range. As an example, it was observed that intrinsic SiO$_2$ of high purity remains uncharged upon radiation, while metal oxide doped SiO$_2$ samples acquire significant charges, the polarity of which depends on the type of metal oxide dopant, which can be explained by trapping of electrons or holes by impurity centers. \cite{miyake2005}

Lacking specific cartilage related data, we find it appropriate to estimate the characteristic charge densities and electric potentials (field) expected under electron radiation of thin films (cartilages). We will assume the following realistic parameters for 6 MeV radiation therapy linac electron beam: electron fluency $\Phi=10^9$ cm$^{-2}$s$^{-1}$, stopping power $w=2$ MeV/cm, cartilage thickness $l=1$ mm and electron-hole pair generation energy $\zeta\sim 10$ eV. We then estimate the charge generation rate per area,
\begin{equation}\label{eq:chargerate}\dot{\lambda}= wl\Phi /\zeta\sim 10^{14} {\rm cm}^{-2}{\rm s}^{-1}.\end{equation}

Under steady state condition, the latter rate must be equal the outflow current density, calculated as $J=(\lambda /l)(2\pi e\lambda)(D/kT)$. Here (in Gaussian units) $\lambda/l$ represents the charge carrier concentration, $2\pi\lambda e/\epsilon$ is the average electric field strength due to the charge density $e\lambda /l$ where $e$ is the electron charge and $\epsilon \sim 10$ is the dielectric permittivity. The electric force is estimated as $F=2\pi\lambda e^2/\epsilon$. The charge drift velocity is $v=F\mu $ and the mobility $\mu$ is related to the diffusivity through the Einstein formula $\mu = D_e/k_BT$. Here $D_e$ is the diffusivity of excess charge carriers. The latter can be either ions or electrons solvated in a liquid water component (unless trapped by ions). The solvated electron diffusivities are on the order of $D_e\sim 10^{-5}$ cm$^2$/s. \cite{barnett1990}.  In summary, the steady state surface charge density and its related electric potential $V\sim Fl/e$ are estimated as
\begin{equation}\label{eq:lambda}
\lambda = \sqrt{\frac{wl^2\Phi k_BT\epsilon}{4\pi \zeta D_ee^2}}\quad {\rm and }\quad V=\sqrt{\frac{4\pi wl^4\Phi k_BT}{\zeta D_e}}\end{equation}

Using the above numerical parameters, Eq. (\ref{eq:lambda}) yields $\lambda \sim 10^{11}$ cm$^{-2}$ and $V\sim 1000$ V.  Because $eV\gg k_BT$, the latter result suggests that the charge build-up creates the electric field which can force ionic drift through the cartilage potentially healing its damages.
The characteristic time $t_c$ of accumulating the charge from Eq. (\ref{eq:lambda}) is estimated from $l=vt_c$ with $v=\mu F$, predicting a rather short process,
\begin{equation}\label{eq:time}t_c=lk_BT\epsilon /4\pi\lambda e^2D_e \sim 1 \ {\rm s}.\end{equation}

When the above charging is due to electron trapping, then the time of its discharge (de-trapping) can be longer by the factor of $\exp(E_b/k_BT)\gg 1$ where $E_b$ is the binding energy of a trapped electron. Therefore, a short pulse of radiation can serve as a trigger for a much longer process bringing in chemical species of desirable polarity into cartilage. This type of treatments becomes possible by injecting certain (trapping) dopants in a cartilage matrix thus constituting a basis for chemo-radiation cartilage treatments.

We shall end this section by noting that the electric field effects on OA cartilages were studied in a number of references. \cite{wang2004,brighton2006,brighton2008,brighton2013,vinhas2020,noruzi2022,zimmerman2024}  Overall, it was found that electric fields of amplitudes from $\sim 20$ mV/cm to $\sim 1$ kV/cm can lead to up-regulation in cartilage matrix, in particular to chondrocyte production, and affect the matrix metabolism. The field application techniques studied \cite{wang2004,brighton2006,brighton2008,brighton2013,vinhas2020,noruzi2022,zimmerman2024} often relied on capacitive coupling assuming minuscule fields of $20 $ mV/cm, unlike here provided estimate of direct charging under radiation producing fields of $\sim 1$ kV/cm [see Eq. (\ref{eq:lambda})]. Other published examples of electrostimulation \cite{chen2020,li2020,li2022} through direct coupling used fields of 1 kV/cm consistent with our estimate. The described electrostimulations used AC electric fields of frequency ranging from 0.1 Hz to 60 kHz. From this manuscript perspectives, the role of frequency is that it makes ions drifting in opposite directions dictated by the field polarity. Such a stirring can be beneficial by supplying ions of opposite signs. In linac-based LDRT implementation, such temporal variations of polarities can be achieved via linac's pulsed beam with axial rotation of gantry adding spatial variation.

\section{Discussion and conclusions}\label{sec:concl}
Various clinical ionizing radiation sources with sufficient penetration depth (keV or MeV energy photons, MeV electrons) appear to be conceivable modalities of OA LDRT. While their planning and delivery details can be quite different, the expected treatment time $\tau$ is a key consideration. Estimated as $\tau =\tau_D\equiv l^2/D$ with the 10 orders of magnitude RED amplified diffusion coefficient, say $\sim 10^4$ cm$^2$/s,  predicts rather short $\tau\sim 1$ $\mu$s. However, such a short time will characterize only the delivery of various species across the cartilage thickness, after which the processes of subsequent cartilage healing micro-structural transformations can last much longer time $\tau = \tau _T\gg \tau _D$ thus forming a significant bottleneck.

The latter inequality enables  self-healing process with much lower diffusivities (than the above discussed many orders of magnitude RED) thus decreasing the dose rate of ionizing radiation. In other words, the emphasis shifts from the fast delivery to a much slower diffusion times providing continuous supply of materials over the characteristic healing times $\tau _T$. The equality $\tau _D\sim \tau _T$ defines the `acceptable' range of diffusivities $D\gtrsim l^2/\tau _T\sim 10^{-6}$ cm$^2$s$^{-1}$ perhaps only moderately exceeding the lowest diffusivities among cartilage species.

The latter consideration points at the radiation induced charge build-up as a natural means of creating long-lasting transport mode described in Section \ref{sec:chargebuild}. As explained in that section, there are impurity centers that are responsible for the electric charge build-up, which suggests a carefully chosen chemical doping of cartilages (perhaps by direct injection) prior to LDRT. The dopants must be selected from those known for their ability to maintain a certain charge state for long times. A list of such elements determined from studying chalcogenide glasses \cite{heo2014} includes mostly rare-earths, such as ions Er3+, Tm3+, Dy3+ or metals like \cite{frumar2003} Ag and Pb. Molecular doping established for proteins \cite{nandi2023} can serve as another choice of charge trapping centers.

As a related curious and yet conceivable effect, we would like to mention that in some patients, individual features of chemical composition can provide charge trapping centers even without intentional doping. For such patients, LDRT is expected to work in a most efficient way. That expectation correlates with anecdotal evidence that some patients claim their arthritis relief in response to very weak  exposure to x-rays during the screening testing.

In conclusion, we have brought in some physics  concepts to the appealing field of the radiation treatments of arthritis. In particular, we have proposed the effect of radiation enhanced diffusion (RED) as a mechanism behind the observed positive effects of radiation treatment of arthritis. In addition, we have identified the effect of radiation induced electric charge build-up as a means of creating long-lasting diffusion flux of species in cartilages, potentially resulting in curative rather than palliative endpoint. Asserting a physical mechanism of curative effect of LDRT depends on developing an objective imaging test complimentary to patient reported outcomes. All in all, our consideration calls upon further research and experimental verifications in the physics of radiation treatment of cartilages.


\end{document}